**The use of next-generation sequencing in personalized medicine**

Liya Popova[1] and Valerie J. Carabetta[1]*

[1]Department of Biomedical Sciences, Cooper Medical School of Rowan University, Camden NJ, 08103

*Correspondence: carabetta@rowan.edu. Tel: +1-856-956-2736

**Abstract**

The revolutionary progress in development of next-generation sequencing (NGS) technologies has made it possible to deliver accurate genomic information in a timely manner. Over the past several years, NGS has transformed biomedical and clinical research and found its application in the field of personalized medicine. Here we discuss the rise of personalized medicine and the history of NGS. We discuss current applications and uses of NGS in medicine, including infectious diseases, oncology, genomic medicine, and dermatology. We provide a brief discussion of selected studies where NGS was used to respond to wide variety of questions in biomedical research and clinical medicine. Finally, we discuss the challenges of implementing NGS into routine clinical use.

**1. Introduction**

*1.1 The rise of genomic medicine*

The variability of individual genomes is responsible for different phenotypic traits, but it also can be more complex in determining predisposition to certain diseases [1]. According to the National Human Genome Research Institute, "genomic medicine" is a fast-growing field that involves the use of an individual's genomic information as part of their care, such as for diagnostic or therapeutic decision-making. From a genetics perspective, the majority of diseases have genetic components and can be divided into three categories: 1. chromosomal, with deletion or addition of entire chromosomes, 2. single-gene (Mendelian) diseases, inherited as familial autosomal or X-linked traits, and 3. complex diseases, which are the results of interactions between certain genetic traits and environmental factors [2]. Since the completion of the Human Genome Project, there has been shift toward pharmacogenomics as a key part of personalized medicine to differentiate responders and non-responders to certain treatments. Integration of genomic medicine sets foot into new stage of clinical medicine implementing genomic-based diagnosis and personalized assessment and treatment [3]. Nowadays, more and more physicians use genetic testing applying to NGS approach, which allows to predict disease predisposition, prognosis, and susceptibility to treatment. Advances in sequencing technology allowed to markedly decrease costs of genetic testing up from three billion dollars costs of Human Genome Project in 2004. Today, whole exome and genome sequencing are performed within a clinically useful timeframe and acceptable costs, making the transition from clinical research trials to availability as a standard clinical tests, outside of global genomic study initiatives [4]. One important application of sequencing technology is in the cancer genomics field and has led to the identification of previously unknown pathogenic mutations. For example, mutations in *BMPR1A* and *SMAD4* have now been linked to colon cancer, and *BRCA1* and *BRCA2* mutations were shown to be associated with early breast and ovarian cancer. These cancers can be prevented or managed by timely detection of mutations and identification of common genetic variants. This allows for the stratification into high and low risk categories for cancer development and the implementation of appropriate clinical interventions, such as early screening for colon cancer, mastectomy, and oophorectomy [5].

Fast development of gene therapy and genome editing platforms during the past decade have given hope for the treatment of some cancers and Mendelian disorders. For example, chimeric antigen receptor (CAR)-T cell immunotherapy is based on genetic engineering of T-cells, has demonstrated efficacy in treating hematologic malignancies and solid tumors, such as B-cell leukemia and lymphoma [6]. Another example is the two most common monogenic disorders, sickle cell disease and β-thalassemia, which could potentially be cured by autologous hematopoietic stem cell transplantation after genetic modification by increasing the expression of non-sickling β-globin or anti-sickling β-like globins. This would lead to restoration of functional, nonpathogenic hemoglobin-containing red blood cells [7]. The recent approval of cystic fibrosis transmembrane conductance regulator (*CFTR*) mutation correctors, such as combinations of ivacaftor, lumacaftor, elexacaftor, and tezacaftor, by the Food and Drug Administration (FDA) for the treatment of cystic fibrosis in children with homozygous F508del mutations is an exciting example of gene therapy transitioning from an idea to extensive research to full clinical application. This class of medicines works stabilizing the mutated CFTR, thereby increasing production of specific ion channels and receptors, improving the lung function by decreasing the build-up of thick mucus. The F508del mutation is the most common, and this treatment can potentially help 70% of patients suffering from cystic fibrosis [8].

Undoubtedly, the development of genomic medicine will continue to impact clinical medicine as the technology advances, setting high hopes for future therapies and possible cures. It is already clear that to reach its full potential, we need to expand the investment in basic science and foundational research. Next generation sequencing (NGS) and genome editing has already led to some major breakthroughs, but there remains a lot of work to be done in the identification of every gene and their variants function and understanding their role in development of disease [5].

*1.2 A brief history of NGS*

The development of sequencing technology began in the late 1970s, with the introduction of two techniques. Maxam-Gilbert sequencing, also known as the partial chemical degradation method, involves chemical cleavage of DNA molecules that are resolved by electrophoresis and bands are visualize using radioactivity [9]. The other, more popular technique is Sanger sequencing. This approach is based on DNA polymerase inserting chain-terminating dideoxynucleotides (ddNTPs) into the growing chain. Incorporation of a ddNTP stops the elongation reaction, resulting in the formation of various lengths extension products, which can be detected by electrophoresis [10,11]. However, significant technological advances occurred rapidly since then and revolutionized Sanger sequencing. The Sanger method was further improved by integration of polymerase chain reaction (PCR) and the use of fluorophores and capillary gel electrophoresis. These techniques were used for the first sequence of the human genome, completed in 2003 as part of the human genome project [12].

The era of NGS, also called next-generation or $2^{nd}$-generation sequencing, began in 2005 with a pyrosequencing method that allowed the achievement of a 100-fold increase in throughput (Figure 1). Pyrosequencing is a technique where a DNA fragment is attached to a bead in an emulsion and luminescence is used to detect pyrophosphate ($PP_i$) release as a nucleotide is incorporated [13]. This sequencing by synthesis (SBS) platform was commercialized by 454 Life Sciences (Roche), but production was discontinued only 10 years later [12,14]. A variation on pyrosequencing was created by Ion Torrent Systems (Life Technologies), which produced a Personal Genome Machine (PGM) sequencer that detects protons released during the reactions (pH changes) instead of fluorescence. This technique allowed for a significant reduction in costs and increased speed of sequencing [15]. In 2006, Solexa (Illumina) produced the Genome Analyzer, a new platform based on cyclic-reversible termination of the polymerization reaction by use of fluorescently labeled 3'-O-azidomethyl-dNTPs. The reaction is "cyclic-reversible" because during the reaction, it temporarily terminates, allowing the removal of unincorporated bases, so that the added nucleotides yield a fluorescent signal [15]. One year later, the sequencing by ligation (SBL) platform was launched by Applied Biosystems (Life Technologies), known as the SOLiD (Small Oligonucleotide Ligation and Detection) system. As the name implies, the ligase-mediated technique determines sequence based on ligation, not nucleotide addition [12,16]. These technologies paved the way for further innovations and improvements that could ultimately lead to personalized medicine in clinical practice.

The next significant breakthrough in NGS was single-molecule real-time (SMRT) sequencing, also known as third-generation sequencing, from Pacific Biosciences [17],. Although single-molecule sequencing introduces a possible difficulty in detecting a signal from the reaction, SMRT can achieve high precision, with a medial accuracy of 99.3% after multiple reads [18]. The downside of single-pass sequence reads is the relatively high error rate (11%-15%), which can be reduced by a sufficient number of sequencing passes to achieve >99% accuracy, or by combining different continuous long reads from the same location [18-20]. The principle behind the SMRT technique is based on the usage of a strand-displacing DNA polymerase, which allows sequencing of the same DNA molecule repeatedly [15,21,22]. Another commonly used SMRT platform is the MinION, released by Oxford Nanopore Technologies. The approach behind the nanopore technology is based on translocation of a single-stranded DNA molecule through a small membrane-bound pore by a molecular machine, like a helicase. During translocation, the electrical potential across the membrane is detected. The MinION is a portable sequencer that allows the user to generate high-accuracy sequencing data in a short period of time [15,23,24]. These advances in DNA sequencing technology led to an exponential growth in performance and increase in cost-efficiency, which enabled the revolution in genomic medicine and has made the possibility of personalized medicine achievable.

In recent years, exponential advances in sequencing technologies, accessibility, and decreasing costs have led to the beginning of large-scale projects in personalized medicine and translational biomedical research of clinical genomics (Figure 2). As a tool to detect different biomarkers, NGS has possible applications in most areas of personalized medicine, from infectious diseases to dermatology. Here, we will discuss the exciting aspects of the multidisciplinary applications of NGS technology in personalized medicine.

## 2. The use of NGS for the management of infectious diseases

*2.1 NGS as a diagnostic tool*

The failure of current diagnostic tests to detect the etiology of infectious processes is not uncommon [25-29]. Timely diagnosis and treatment of a pathogen can be crucial for survival and prevention of transmission. Emerging pathogens represent a large threat to public health in the era of globalization and rapid migration between countries and continents. The application of NGS allows for accurate detection of both existing and novel pathogens in a timely manner. For example, NGS technologies were used to characterize the genomes and for the detection of Bas-Congo rhabdovirus, MERS-CoV and SARS-CoV-2. Bas-Congo rhabdovirus was responsible for the outbreak of acute hemorrhagic fever in Africa in 2009, with the mortality rate often exceeding 50% [30,31]. Middle East Respiratory Syndrome coronavirus (MERS-CoV) first emerged in 2012 in Saudi Arabia with a reported fatality rate of 32.7% [32]. Timely identification of the infectious agent is a prerequisite to identifying or developing potential treatment options. This was exemplified with the response to SARS-CoV-2, the causative agent of ongoing COVID-19 pandemic that is responsible for over 6.9 million deaths worldwide as of April 2023, according to the World Health Organization (WHO, [33]). NGS allowed for the timely sequencing of the SARS-CoV-2 genome and this information was rapidly disseminated across research laboratories around the globe, thereby beginning and leading to the unprecedented development of multiple effective vaccines in a short period of time. In fact, NGS continues to play an important role in identification of new variants of SARS-CoV-2, early detection, and prevention of the spread [34]. The rapid turnaround and cost effectiveness of NGS has made it possible to rapidly respond to global emergencies.

*2.2 NGS contributions to epidemiology of infectious agents*

NGS is also beneficial towards the management of infectious diseases on a smaller scale. Practical application of high-resolution sequencing for clinical epidemiology has allowed for deep analysis of mutants, transmission pathways, and evolution of bacterial populations. For example, NGS was used to study outbreaks of *Staphylococcus aureus* in hospital settings and its spread to the community [35]. In one study of an outbreak of methicillin-resistant *S. aureus* (MRSA) in the neonatal intensive care unit, whole-genome sequencing of 14 isolates and subsequent reconstruction of a phylogenic tree revealed that outbreak-related isolates were resistant to gentamicin and mupirocin. This was unusual for MRSA strains across the United Kingdom and differed from the typical antibiogram pattern of resistance to cefoxitin, erythromycin, and ciprofloxacin. Köser et al. performed whole genome sequencing to create an evolutionary tree of outbreak and non-outbreak strains, which revealed clusters based on antibiotic resistance and toxins [36]. This information could have been used to inform treatment decisions and contain the outbreak faster.

In 2010 in Haiti, whole-genome sequencing was used to detect the source of an epidemic of *V. cholerae* O1, given that there was no documented evidence of any previous outbreaks on the island. Katz et al. used NGS to discover that the strains derived from a common ancestor, and a later epidemiological analysis showed that the outbreak originated from Nepal, where United Nations peacekeepers were working in unsanitary conditions [37]. Katz et al. further demonstrated that only *de novo* mutations were responsible for genetic polymorphisms that led to drug resistance in these clinical isolates, without a major contribution from horizontal gene transfer [38]. However, this is not a universal observation, as NGS analysis of *V. cholerae* strains originating from the Queensland waterway in Australia demonstrated that the waterway serves as a reservoir for virulence genes that can be transferred rapidly between highly-divergent strains via mobile genetic elements [39]. These studies demonstrate two powerful uses of NGS, in tracing the origins of outbreaks and in the understanding of how virulence genes spread among populations.

There was an outbreak of *Escherichia coli* O104:H4 that spread throughout Germany on contaminated bean sprouts in 2011. One of the clinical features of the infection was colonic ischemia, which is unusual for this type of infection, in addition to bloody diarrhea, with the complication of hemolytic-uremic syndrome. NGS allowed for the quick identification of the outbreak strain, which was enteroaggregative *E. coli* that had acquired Shiga-toxin 2 and antibiotic resistance genes that were responsible for the unusual clinical presentation [40]. A final important application of NGS towards our understanding of epidemiology of antibiotic resistant species was in agriculture. NGS was used to demonstrate that the use of antibiotics in farm animals resulted in the occurrence of antibiotic-resistant strains. Furthermore, it was found that animal and poultry husbandry can be a potential reservoir for antibiotic-resistant human pathogens [41,42], which was one of the first demonstrations of this phenomenon. Clearly, NGS has made valuable contributions to our understanding of epidemiology and its use will only continue to grow moving forward.

*2.3 NGS and the evolution of drug resistance*

NGS is powerful for the study of selection targets or the dynamics of genome evolution that occur in a population over time. Applications for the study of evolution might include genome-wide association studies (GWAS) or transposon insertion sequencing (Tn-seq, [43,44]). Whole-genome sequencing of *S. aureus* clinical isolates has aided in our understanding of the mutations that accumulate to drive the evolution of drug resistance *in vivo* [45]. In one study of cystic fibrosis patients, GWAS analysis revealed a high frequency of mutations in the alternate sigma factor SigB, isolated from *S. aureus* strains from chronically infected patient airways [46]. There seems to be an *in vivo* selective pressure to attenuate SigB activity, which is interesting because SigB regulates the general stress response and expression of virulence factors, like capsule formation and alfa-toxin production [47,48]. Interestingly, *sigB* mutants poorly formed persisters, which are slow growing, drug-tolerant cells that exist in a population of genetically identical cells, when compared to clinical strains. Therefore, while examining the evolution of drug resistance, the NGS data also potentially opened the door for the design of new antibacterial therapeutics to control refractory infections [48]. In *Vibrio cholera,* the occurrence of strains resistant to commonly used antibiotics remains a large problem. Using NGS to study clinical isolates of *V. cholerae* in Laos, Iwanaga et al. determined the mechanism of antibiotic resistance. They showed that the collection of *V. cholerae* strains acquired the SXT constin element, an antibiotic resistance cassette encoding resistance to chloramphenicol, tetracycline, streptomycin, and sulfamethoxazole [49,50]. With this information, it may be valuable to screen isolates by PCR for the presence of the SXT constin element, which could inform treatment strategy.

One controversial question about the acquisition of drug-resistance genes by *Mycobacterium tuberculosis* was answered using NGS. For a long time, it was believed that drug resistance can only emerge in patients with active tuberculosis (TB) disease, while bacteria in the latent state have little capacity for mutation since they are dormant or slowly growing. People with latent TB are infected with *M. tuberculosis*, but they do not have clinical symptoms and therefore cannot spread the disease to others. Someone with active TB disease has clinical symptoms and can spread *M. tuberculosis* to others. Because of this, latent infections are typically treated with one first-line antibiotic, isoniazid. The use of high-density, whole-genome sequencing allowed for the assessment of the capability of *M. tuberculosis* to acquire mutations and evolve over the course of infection. It was found that *M. tuberculosis* acquires mutations during latency despite the markedly low replication rate. Surprisingly, mutation rate and the number of chromosomal mutations observed during latency was like that observed in *M. tuberculosis* in the active state. This finding suggests that monotherapy for latent tuberculosis is a risk factor for emergence of resistance and there is need to reconsider how latent TB infections are treated [51]. The use of NGS can improve and directly influence physician treatment choices.

2.4 Metagenomics

Another major challenge in management of infectious diseases is the detection of etiological pathogens from complex, polymicrobial patient specimens. *In vivo*, bacteria live in multispecies communities, and identifying the clinically relevant species is difficult. This is because standard laboratory growth techniques are not always sufficient for the cultivation of all the microbial species from a specimen. Clinically significant organisms might be slow-growing or get out-competed by other strains when co-cultured under artificial laboratory conditions [52]. The use of a metagenomic analysis allows for the analysis of these communities without prior culturing, which also improves turnaround time for results. It has been shown that deep sequencing of 16S rRNA, with appropriate downstream bioinformatic analysis yields high accuracy data and allows for the detection of bacterial species in heterogeneous clinical specimens that cannot be identified using conventional clinical laboratory methods [53]. This approach for the recognition of novel, disease-causing species has the potential to become a diagnostic tool for use in clinical practice. In addition, this type of analysis could be used for epidemiologic surveys and improvement of sequence databases. The largest advantage of metagenomic analysis is the rapid turnaround, as it removes the 16-24 hours required for culturing, and it is relatively inexpensive, making it ideal for routine clinical use.

Metagenomics have also been used for the characterization of the human microbiome. The normal human microbiome is diverse and represents symbiotic relationships between host and microbes, like bacteria and fungi, that form a complex community. Disruption of normal microbiota has been linked to many diseases, including depression, Alzheimer's disease, autoimmune diseases, and coagulation disorders [54]. Metagenomic analysis has been widely applied to determine the composition and diversity of the gut, vaginal, oral, skin, and placental microbiome in health and disease (reviewed in [55-57]). Other valuable information gained from this type of study is the determination of essential regulators or factors that are necessary for interspecies or host interactions, and the distribution of antibiotic resistance genes or virulence genes among populations [55]. As soon as the essential regulatory genes and factors are identified, it will open the door for potential biomarker screening or therapeutic development. For example, these could lead to the identification of novel, pathogen-specific bacteriophages that could be used for the treatment of antibiotic-resistant bacterial pathogens [58,59].

Periodontitis is a polymicrobial biofilm-induced disease that leads to dysbiosis of the oral microbiome, then to inflammation and bone loss [60,61]. In order to determine the underlying mechanisms of dysbiosis, Yost et al. characterized gene expression patterns using a meta-transcriptomic analysis of the entire oral microbiome during the progression of periodontitis [62]. In samples with periodontal disease progression, analysis demonstrated changes in gene expression in major periodontal pathogens, including the upregulation of genes associated with transport, proteolysis, and the protein kinase C-activating G-protein coupled receptor signaling pathway. Comparison of follow-up non-progressive periodontal disease and baseline samples did not identify any differentially expressed genes, implying that clinically stable sites do not have changes in microbial gene expression. Additionally, specific pathogens associated with periodontitis showed individual patterns of genes expression. For example flagellar biosynthesis genes were upregulated in *Treponema denticola*, and biotin synthesis genes and capsular polysaccharide biosynthesis were upregulated in *Porphyromonas gingivalis* [62]. This is one example of how metagenomics in combination with other NGS technologies could become a powerful tool to understand complex communication and relationships among polymicrobial communities.

*2.5 NGS in vaccine design*

One of the most important applications of NGS in clinical research night be new vaccine design. Most vaccines that helped to combat the severe infectious diseases of 20[th] century were developed based on Pasteur's approach from 1885. This involved the isolation of the pathogenic microorganism, followed by inactivation, and finally injection (reviewed in [63]). Many vaccines that are still in use today were developed using this formula, including the killed-inactivated polio vaccine, and live attenuated vaccines against measles, mumps, and rubella (MMR), varicella zoster and rotavirus. In subsequent years, new technologies became available, which led to considerable progress in vaccinology. One breakthrough was the development of recombinant vaccines, where surface antigens are expressed via plasmids in bacterial or yeast hosts for insertion into virus-like particles [64]. This avoided the possibility of reversion of the attenuated strains to wildtype and subsequent infection. The Hepatitis B (HBV), Human papilloma virus (HPV) and pertussis vaccine were all created using this technology.

The next step in the evolution of vaccine development is reverse vaccinology, or genome-based antigen discovery. This approach is largely based on data from NGS analysis of the pathogen of interest's genome, where the sequencing of genomes from a large number of strains, serotypes, or related species allows for the identification of broadly distributed, potential antigens for vaccine development. One benefit of NGS was the discovery of new antigen determinants in pathogens that can avoid host immune responses, such as *Streptococcus pneumoniae*, *Streptococcus agalactiae, S. aureus*, and *Chlamydia pneumoniae* [65,66]. While promising antigens have been identified, these have not been developed into effective vaccines for practical use. Even though reverse vaccinology may be useful for some pathogens, its potential is limited for pathogens with highly variable surface antigens, like human immunodeficiency virus (HIV) or respiratory syncytial virus (RSV), for which structural vaccinology is a more promising approach [63]. Currently, the evolution of vaccinology has led to synthetic biology, that uses *in vitro* synthesis of DNA and RNA for vaccine development. One of the most well-known examples of synthetic biology is the mRNA vaccines developed against SARS-CoV-2. NGS was used to analyze the genome and transcriptome of the virus. The coding sequence of a specific spike protein was selected as a vaccine candidate. The target gene was inserted into plasmid, transcribed into an mRNA molecule, and packaged into liposomes to protect the mRNA [67,68]. The success of these vaccines provides strong evidence that NGS can be used for future vaccine development.

**3. NGS in genomic medicine**

*3.1 The use of NGS to study single gene disorders*

One of the most exciting clinical applications of NGS is for genomic medicine. Although it is a relatively new discipline, genomic medicine is already transforming therapeutic approaches and the management of patients with genetic diseases, especially single gene, Mendelian disorders, and certain cancers. Rapid testing of all genes related to a disease is particularly useful in patients with rare diseases, such as hereditary ataxia, amyotrophic lateral sclerosis [69], an unusual clinical course, or an atypical response to treatment [70]. In addition, for cases where whole exome sequencing (WES) is not successful or accurate, RNA-seq is being explored as an alternative diagnostic strategy for simple Mendelian disorders [71]. Since 2012, a global initiative brought together major institutions to support the opening of Centers for Mendelian Genomics, where the application of NGS technologies has been used for gene discovery [72]. For example, WES was used to identify a mutation in the dihydroorotate dehydrogenase (*DHODH*) gene, which is responsible for postaxial acrofacial dysostosis, also known as Miller syndrome [73]. Interestingly, this approach led to the discovery of a coincidental mutation in dynein axonemal heavy chain 5 (*DNAH5*) gene, which is associated with primary ciliary dyskinesia, in two patients with Miller syndrome that provided the underlying cause of their recurrent respiratory tract infections. The identification of gene mutations in poorly understood diseases will allow for improved diagnostics and improved care [74].

One of the most successful applications of the use of a DNA-based sequencing approach in genomic medicine was for inherited platelet diseases. The first worldwide studies in this field were initiated by BRIDGE (Bleeding and Platelet Disorder consortium) in 2011, where NGS was used for inherited platelet diseases to create a gene platform based on whole exome and whole genome sequencing of patients with the disease [75]. Since then, NGS has been extensively used to establish the diagnosis, discover new alleles resulting in inherited platelet disorders and for genetic counseling [76] . For example, the use of whole genome sequencing identified new variants of diaphanous–related formin 1 gene (*DIAPH1*) that lead to sensorineural hearing loss-thrombocytopenia syndrome [77]. In another study, Gorski et al. used WES to uncover novel genetic variants that were responsible for the development of neutralizing antibodies against factor VIII in patients with severe hemophilia A. Factor VIII is commonly used as a replacement therapy in such patients. 17 variants were identified that were associated with a high risk of development of neutralizing antibodies and 11 variants were identified as protective. From this work, they identified a possible susceptibility locus for antibody development against factor VIII, in the *LCT* locus, located on chromosome 2q21 [78]. This study demonstrated that NGS is effective for the identification of rare variants in the population and can be used to identify new susceptibility loci for genetic diseases.

*3.2 The use of NGS to study complex traits*

Understanding non-Mendelian, multigene diseases can also benefit from the application of NGS, with recent advances being made in cognitive, neurological, cardiovascular, and psychiatric disorders. To detect genetic variations to diagnose a specific disorder or group of disorders for congenital heart diseases, targeted gene panels have been developed. For example, 13 genes associated with common congenital heart disease are available as a NGS congenital heart disease panel for prenatal testing [79,80]. The majority of congenital heart disease occurs sporadically, but certain environmental factors and genetic disorders have been associated with increased risk [81]. For example, *GATA4* mutations are associated with familial ventricular septal defects [82], while mutation in transcription factor *NKX2-5* was found in families with atrial septal defects and conduction delays [83]. In DiGeorge syndrome, a chromosomal microdeletion of 22q11 results in the loss of T-box transcription factor *TBX1*, which is responsible for the associated aortic arch defects and is associated with majority of cases [84]. Although NGS should be a powerful tool for the detection and diagnosis of inherited cardiac disease, obstacles arose when trying to apply NGS to detect disease alleles in sporadic cases. This because large-scale screening on the population level revealed a high quantity of potentially pathogenic variations in healthy individuals [85]. Thus, large population linkage studies are required to determine the contribution that these alleles play in development of cardiac disease. Others propose that target gene panels remain the most practical for use in a clinical setting, but WES has great value in the research environment for novel gene discovery and the identification of rare variants [86,87]. WES and RNA-seq was recently applied to examine rare copy number variants (CNVs), which are associated with congenital heart defects. From this study from approximately 2400 families, association of CNVs with specific sub-types of congenital heart disease were identified and five novel CNVs were identified [88]. Clearly, NGS analysis will continue to be important for our foundational knowledge of congenital heart disease, even if its use is limited for routine clinical diagnostics.

*3.3 Applications of NGS for neonates and pre-neonatal diagnostics*

One of the exciting possible clinical applications of NGS is in neonatal and pediatric medicine to establish diagnosis of genetic diseases in a timely manner. This is especially important where a delayed or phenotype-based clinical decision may lead to a substantial increase in morbidity or mortality [89,90]. One children's hospital study in neonatal and pediatric intensive care units (ICU) analyzed the diagnostic rate of infants younger than 4 months old with suspected genetic disorders and congenital anomalies. Rapid whole-genome sequencing, which could be completed in 50 hours, was compared to standard genetic testing techniques, like comparative genomic hybridization arrays and fluorescence *in situ* hybridization (FISH). 20 out of the 35 critically ill infants were diagnosed with a genetic disease by rapid whole-genome sequencing and three out of 32 by standard genetic testing. 18 of those 20 diagnoses established by whole-genome sequencing were not identified with standard genetic testing, and nine of them were diseases that were not considered on the original differential diagnosis list. Furthermore, 13 cases led to a modification of the acute clinical management, including palliative care, medications, and genetic counselling [91]. This clearly highlights the need for the rapid, accurate diagnosis of genetic diseases, as it can lead to improved management and patient outcomes. In another single hospital study,

WES was performed for 278 infants in the first their 100 days of life that were hospitalized in the neonatal ICU (NICU), with suspected genetic diseases based on clinical indications. 102 infants met the criteria for molecular diagnosis and approximately 39 of them had an atypical or unrecognized clinical presentation. Establishment of the molecular diagnoses altered subsequent medical care, leading to medication changes, transplant consideration, and initiated new subspecialist care, which was not contemplated before the NGS analysis [92]. The use of NGS for the diagnosis and management of genetic diseases among neonates will likely become widespread in the near future.

Another revolutionary application of NGS is in its use for prenatal diagnostics. The discovery of circulating cell-free fetal DNA in the plasma of pregnant women made it possible to detect paternally inherited alleles by selective, targeting PCR analysis. However, this methodology has not been accurate for detecting fetal genetic aberrations or maternally inherited genetic disorders [93]. The implementation of NGS allowed for screening for multiple mutations in a non-invasive way. This includes the difficult-to-detect fetal autosomal recessive, monogenic disorders, and aneuploidies. With the increasing affordability and a rapid turnaround time, the possibility of instituting non-invasive prenatal screening using NGS technologies as a common screening test in the routine prenatal clinical practice is becoming more realistic [94]. In fact, NGS is being applied to diagnose fetuses with congenital heart defects [95]. However, technical challenges likely remain, such as the difficulty in distinguishing fetal and maternal genomes. In addition, there are ethical considerations and the establishment of clear guidelines for testing and the implementation of genetic counseling before and after the testing are required.

**4. NGS in oncology**

*4.1: NGS for cancer diagnostics*

NGS technologies can promote advancements in the field of cancer research, including uncovering fundamental knowledge about the complex mechanisms of pathogenesis and fostering new therapeutics.

NGS can detect genetic abnormalities at the single gene level, changes in gene expression and facilitate the molecular description of tumors. Together, this information could lead to revision of the current approaches in establishing diagnosis, treatment, and prognosis of neoplastic diseases [96,97]. Over the past few decades with the application of newer NGS technologies, there have been many important breakthroughs. One such breakthrough was the revision of breast cancer diagnosis and management based on genetic subtypes. Specific genetic abnormalities have been associated with different outcomes, and whole genome sequencing and transcriptomic analyses prompted the reevaluation of the risk of relapse and response to treatments [98]. This led to the discovery of breast cancer molecular subtypes that benefit from adjuvant chemotherapy. For example, oestrogen receptor-$\alpha$ (ER$\alpha$) is a ligand-dependent transcription factor regulates cell growth genes and plays an important role in breast cancer progression. More than 70% of breast cancer cells express ER$\alpha$ which is encoded by the gene *ESR1* [99]. Recently, chromatin immunoprecipitation followed by NGS (ChIP-seq) was used by Ross-Innes et al. to determine ER-binding sites in tumors from patients with different clinical outcomes This study revealed that in patients with Tamoxifen-resistant (a drug used to treat hormone receptor-positive breast cancer) breast cancers, there was a unique pattern of ER-binding sites compared to drug-susceptible cancer cells [100]. ChIP-seq analysis of clinical samples showed that tamoxifen resistance is related to ER binding affinity, with differential ER binding genome profiles associated with distinct clinical outcomes. For patients with good clinical outcomes, there were ~500 ER binding events in the genome and in patients with poor outcomes and metastases, there were 1192 events, representing the acquisition of additional, tumor-specific ER-binding regions. This is one of the seminal studies that detected the association between ER-binding in ER+ breast cancers and clinical outcomes [100]. In addition, this analysis led to the identification of a new resistance mechanism in ER+ breast cancer cells, which could become a screening tool to determine if tamoxifen will be an effective chemotherapy.

One of the few solid cancers that is difficult to analyze by NGS is melanoma, due to a high number of mutations from ultraviolet light. Hodis et al. performed whole genome sequencing and identified novel candidate genes (*SNX31, PPP6C, TACC1, STK19, RAC1, ARID2*) by comparing baseline mutation level to mutation in genes of interest [101]. WES analysis of 147 tumors revealed that 12% had mutations in *PPP6C* in tumors with additional mutations in *BRAF* or *NRAS.* Mutation in *BRAF* is known to be involved in the pathogenesis of metastatic melanoma. This discovery and approval of the BRAF kinase inhibitor Vemurafenib started a new era in personalized therapeutics of melanoma. Application of NGS to identify treatable mutations such as *BRAF* in each patient is a promising approach in personalized treatment that has already been implemented at oncological centers (Figure 3, [102,103]).

*4.2 NGS to study the underlying genetics of cancer and metastasis*

Androgen hormones, whose effects are mediated through androgen receptors (AR), play a significant role in the progression of prostate cancer. Like the ER receptors in breast cancer, mutations of the ARs in early prostate cancer are infrequent but are commonly found in advanced androgen-independent tumors. The abnormal activation of ARs leads to series of intracellular events that initiate transcriptional activity and promote uncontrolled growth. ChIP-seq was used to study AR binding across the genomes of prostate cancer cell lines. Analysis revealed that NKX3-1 is a novel transcription factor that cooperates with ARs to regulate transcription of genes that drive prostate cancer progression [104,105]. Thus, the use of NGS technology can uncover proteins that work together to regulate transcription of growth-promoting genes, which aids in our fundamental understanding and possibly will lead to novel therapeutics.

Despite the advances in surgical treatment and chemotherapeutics, lung cancer is still one of the deadliest cancers, with high lethality due to metastatic disease [106]. Often, the disease is diagnosed in later stages when it becomes symptomatic and already resulted in extrapulmonary metastases. One focus of lung cancer research is genome analysis and gene expression, with the goal of gaining a better understanding of the progression to metastatic disease. Recent studies have shown that microRNAs (miRNAs) have been linked to aberrant gene expression in cancer cells of different origins [107]. miRNAs are short, 22 nt molecules that play important regulatory roles in various fundamental biological processes, like cellular proliferation, apoptosis, and differentiation [108]. So far, three miRNAs (has-miR-183-3p, has-miR-183-5p, has-miR-574-5p) have been linked to lung adenocarcinoma metastasis. NGS and qRT-PCR was used to confirm differential expression of these miRNAs in cancer samples with and without metastases. Expression profiling analysis showed decreased expression of has-miR-183-3p and has-miR-183-5p in metastatic samples of lung adenocarcinoma. Conversely, has-miR-574-5p was increased in metastatic samples. miRNA-seq was used to identify numerous additional miRNAs linked with tumor survival and progression [109].

The expression of other types of small RNAs have also been linked to the presence of distant metastases in lung adenocarcinoma cells. Daugaard et al. performed small RNA sequencing to profile the transcriptomes of lung adenocarcinoma cells [110]. PIWI-interacting RNAs (piRNAs) are 26-31 nt in length and are not well conserved. They are known to be involved in transposon silencing, genome rearrangement, epigenetics, stem-cell maintenance, and tumorigenesis [111,112]. piRNAs and miRNAs both play a significant role in the development of distant lung cancer metastases. For piRNAs, there was a significant association between the decreased expression of piR-57125 and distant metastases. For miRNAs, decreased expression of miR-30a-3p and increased expression of miR-210-3p was linked with distant metastases [110]. Recently, it was shown that tumor cells release miRNAs into blood, where they stay stable [113]. Thus, genome-wide expression analysis of circulating miRNAs and possibly piRNAs from serum can be used as biomarkers for metastatic progression and overall prognosis.

*4.3 Challenges in implicating NGS technology in clinical practice*

The potential of NGS technology in a clinical setting is vast. However, the disputes about the use of NGS in diagnosing, monitoring and prognosis of different cancers are ongoing. The discovery of new cancerous biomarkers because of deep genomic screens provides support for the implementation of NGS into clinical practice, although it would still need extensive validation and approval [114]. The diagnosis, treatment, and management of leukemias or lymphomas is one example where clinical practice would benefit from NGS implementation. When these cancers develop, there usually is a single T or B cell clone that uncontrollably proliferates, which has a unique T- or B-cell receptor sequence. Since this malignant clone can compose more than a half of the entire B- or T-cell population in cancer patients, the clonality of receptors is used for diagnosis of blood cancers and tracking residual disease following treatment [115]. NGS analysis of T- and B-cell receptor genes from blood has been proposed for the assessment of treatment response and risk stratification in patients with acute lymphoblastic leukemia, with minimal residual disease. The use of NGS instead of allele-specific, quantitative real-time PCR would be more informative and accurate in the characterization of the lymphoid population to monitor disease evolution and detect relapse as early as possible [116].

For the diagnosis of non-Hodgkin's lymphoma, a controversial approach was proposed to detect free DNA in plasma as a source of cancer-specific DNA instead of white blood cells. The challenge with diagnosing lymphomas is the absence of known point mutations and difficulty in detecting circulating tumor cells. This proposal is based on the fact that all lymphomas have rearrangements of their immunoglobulin genes. The idea is to use NGS to identify heavy chain immunoglobulin (*IgH*) gene rearrangement by capturing and sequencing the *IgH* genomic regions. With knowledge of the rearranged sequences, circulating fragments DNA can be detected in plasma, thereby working as a tumor progression marker. This approach has the potential for clinical application to diagnose and monitor disease progression, but still requires large-scale clinical trials [117].

In a study of acute myeloid leukemia patients, a NGS panel was tested for accuracy compared to routine, clinically approved methods, such as hybridization probes and fragment analysis by Sanger sequencing of positive samples. A 19-gene NGS panel targeting specific acute myeloid leukemia genes was used for a cohort of 162 patients. Comparison with conventional techniques showed 100% concordance, and NGS was able to detect more clinically relevant mutations. This NGS panel was highly sensitive and specific, with cutoff points of 3% and 5% for point mutations and insertions/deletions, respectively. This panel was proposed for use in diagnostic and prognostic classification [118].

Although NGS appears to be a breakthrough technique in establishing the molecular diagnosis of malignancies, even NGS is not ideal. Whole-genome sequencing provides large amount of data with single nucleotide resolution, but the costs might still be too high for routine clinical implementation in all practices [119]. For example, the cost of NGS in clinical settings in the United States between 2016-2019 were on average between $438 and $3,700 for tumor mutation burden tests and from $1,722 to $2,249 for hereditary cancers [120]. WES is a more cost-effective alternative to whole-genome sequencing, but it is also can be expensive and there is a still a need to analyze large amount of data [119]. The discovery of new biomarkers based on large sequencing data requires the implementation of large gene screening panels, which would be challenging to introduce into routine clinical practice because of the high-volume of tests and slower turnaround time. This would require additional resources, costs, and expertise to interpret high complexity data. Currently only smaller, targeted gene panels are considered practical for tumor classification. For the development of future NGS clinical panels, the role of mutations in oncogenesis, detection of clinically important polymorphisms, translocations or gene fusions, turnaround time, and costs of laboratory testing must all be considered [121,122].

**5. NGS applications in dermatology**

The use of NGS in investigative dermatology started with the application of NGS to determine the underlying immune response associated with specific insults. Gaide et al. used a technique called immunosequencing, which combines multiplexed PCR with NGS to monitor V(D)J recombination of the T-cell receptor gene [123]. Using immunosequencing, they examined the T-cell response after skin vaccination and were able to clarify the clonal origin. It was discovered that resident memory T cells that reside in the peripheral tissue and the central memory T-cells from the lymph nodes have the same T-cell progenitor, which creates memory T cells with different effector properties, but the same antigen specificity in different compartments. Peripheral skin T cells display a rapid hypersensitivity response, while central lymph node T cells are responsible for a delayed response. This data provides the underlying mechanism for allergic contact dermatitis and explains the propensity for recurrence and the refractory properties of this disease [124,123].

NGS of the T-cell receptor has also been used for both diagnostic and prognostic purposes for cutaneous T-cell lymphoma (CTCL, [125]). Detection of a malignant clone is crucial in making the diagnosis of CTC and the widely used diagnostic test of T-cell receptor $\gamma$ PCR only detects specific subsets. It was recently demonstrated that NGS is more specific and sensitive in the detection of malignant T-cells and can accurately distinguish it from benign inflammatory diseases. Another application that NGS was useful for was in treatment response monitoring and identification of disease recurrence. As NGS can be customized for personal medicine, it could be used to monitor individual responses to therapy and detect prognostic or diagnostic biomarkers at any stage of the disease.

## 6 Conclusions

NGS is a powerful research tool that accelerates new discoveries and gives broader opportunities for research in the fields of genomics and personalized medicine. The discussion about the potential application of NGS in the field of personalized medicine and day-to-day clinical practice has been ongoing.

The great excitement arises from possibility to improve diagnostic and prognostic approaches in certain areas of medicine, like genetic disease and oncology. Ultimately, these improvements could lead to earlier diagnosis, earlier detection of relapse, and more favorable clinical outcomes. Perhaps the most successful clinical application of NGS is in oncology, in which the adaptation of diagnostic and therapeutic strategies is based on gene signatures and mutational profiles of individuals [96,97,113]. Despite the technological advances, most disease treatments are based on "standards of care" approaches, which might not work for every patient. In addition, patients that are unresponsive to standard treatments are met in clinical practice quite often, especially in the fields of cancer or rare diseases. Unfortunately, in such clinical cases, rounds of ineffective treatment can lead to delays and unfavorable outcomes before the appropriate treatment has been identified. Implementation of NGS, together with clinical findings and laboratory tests, could help to diagnose individual variations and predict certain predispositions and responsiveness to distinct treatment schemes [126]. For example, the development of cancer targeted therapies became possible after the identification of specific, causative DNA mutations. The first whole-genome sequencing of a cancer patient was performed in 2008. The patient had acute myelogenous leukemia, and WGS revealed 10 genetic mutations that were relevant to the disease, opening the door to genomics as a fundamental weapon in the battle of cancer [127]. Since then, whole-genome sequencing has been expanded to other cancers and led to worldwide cancer sequencing projects. It is not just oncology that has benefited from the implementation of NGS into clinical practice. Establishing a diagnosis of a rare genetic disease has been a challenge for decades, leaving doctors with only a few diagnostic approaches and an absence of tools to find disease variants. NGS overcomes these obstacles, offering greater sensitivity and increasing the likelihood accurate diagnosis of rare diseases, including the detection of different genetic variants. For example, different variations of genetic mutations in neonatal diabetes mellitus will respond differently to available treatments, and NGS allows for their accurate detection [128]. The use of genomic information has pushed the evolution of gene therapy and expanded feasibility of personalized medicine.

Although NGS is an excellent tool and can be extremely beneficial in clinical settings, it is still facing multiple obstacles on the way to routine practice. These obstacles include the clinical laboratory environment, secure storage of large amounts of generated data, and the need for validation of testing platforms on broad populations. Additional future challenges include improving the turnaround time for the delivery of laboratory results, and the need to educate clinicians to analyze the data effectively and accurately. However, the clinical application of NGS should be prioritized because of the therapeutic potential. Knowledge of the genetic causes of a medical condition through NGS can lead to improved therapeutic options, even if it is not a cure. Its adaptation and adjustment into clinical practice requires broader expansion, faster processing of large amounts of data, and a standardized, affordable educational platform for appropriate personnel. Unquestionably, NGS is a powerful tool and its broad implementation into clinical practice will be of benefit to both physicians and patients.

**Acknowledgement:** This work was supported by grant GM138303, awarded to V.J.C.

**Table 1:** Commonly used genes in prenatal testing panels of congenital heart disorders.

| CHD Genes | Disorder | Inheritance Pattern |
|---|---|---|
| CHD7[b] | CHARGE syndrome[b] | AD |
| | Isolated CHD52 | |
| ELN[a] | Supravalvular aortic stenosis | AD |
| GATA4[a] | Atrial septal defect 2 | AD |
| | Atrioventricular septal defect 4 | |
| | Tetralogy of Fallot | |
| | Ventricular septal defect 1 | |
| GATA6[b] | Atrial septal defect 9 | AD |
| | Atrioventricular septal defect 5 | |
| | Pancreatic agenesis and congenital heart defects[b] | |
| | Persistent truncus arteriosus | |
| | Tetralogy of Fallot | |
| GDF1[b] | Right atrial isomerism[b] | AR |
| | Congenital heart defects, multiple types, 6 | AD |

| CHD Genes | Disorder | Inheritance Pattern |
|---|---|---|
| JAG1[b] | Alagille syndrome[b] | AD |
| | Tetralogy of Fallot | |
| NKX2-5[a] | Atrial septal defect 7 | AD |
| | Conotruncal heart malformations, variable | |
| | Hypoplastic left heart syndrome 2 | |
| | Tetralogy of Fallot | |
| | Ventricular septal defect 3 | |
| NKX2-6[a] | Conotruncal heart malformations | AR |
| | Persistent truncus arteriosus | |
| NOTCH1[b] | Aortic valve disease 1 | AD |
| | Adams-Oliver syndrome 5[b] | |
| NR2F2[a] | Congenital heart defects, multiple types, 4 | AD |
| TBX1[b] | DiGeorge syndrome[b] | AD |
| | Velocardiofacial syndrome[b] | |
| | Conotruncal anomaly face syndrome[b] | |
| | Tetralogy of Fallot | |
| TBX5[b] | Holt-Oram syndrome[b] | AD |
| | Atrioventricular septal defect, Atrial septal defect, Ventricular septal defect | |
| ZIC3[b] | Congenital heart defects, nonsyndromic, 1, X-linked | XLR |

| CHD Genes | Disorder | Inheritance Pattern |
|---|---|---|
| | Heterotaxy, visceral, 1, X-linked[b] | |
| | VACTERL association, X-linked[b] | |

AD, autosomal dominant; AR, autosomal recessive; XLR, X-linked recessive; NS-CHD non-syndromic congenital heart disease. Adapted from Ref. [80]. Copyright, The Authors, 2022. Published by KeAi Publishing. Distributed under the terms of the Creative Commons Attribution License (http://creativecommons.org/licenses/by/4.0/).

[a]Gene associated with NS-CHD only.

[b]Gene associated with NS-CHD and extracardiac presentation.

**Figures**

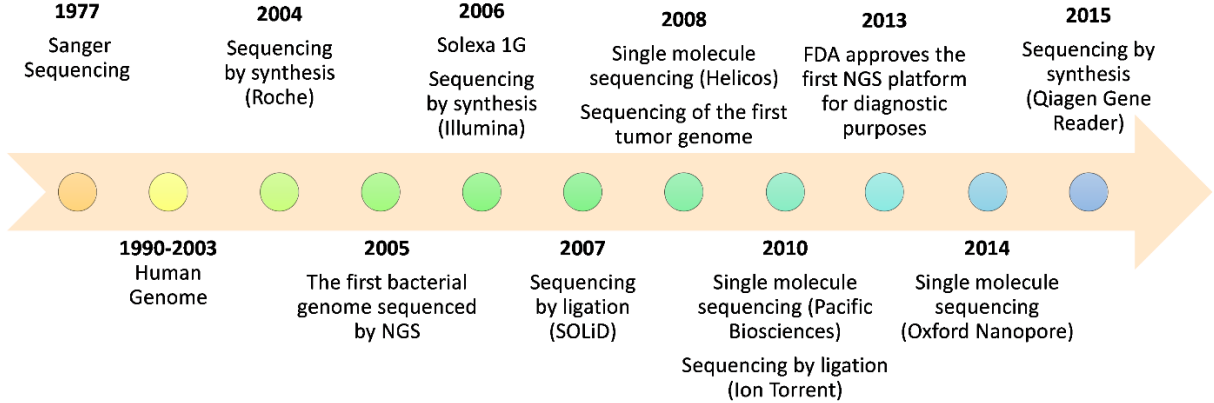

**Figure 1.** Timeline of commercial NGS platforms. A brief history of the instruments introduced to the market between 2004 and 2015 and developmental milestones. NGS; next-generation sequencing.

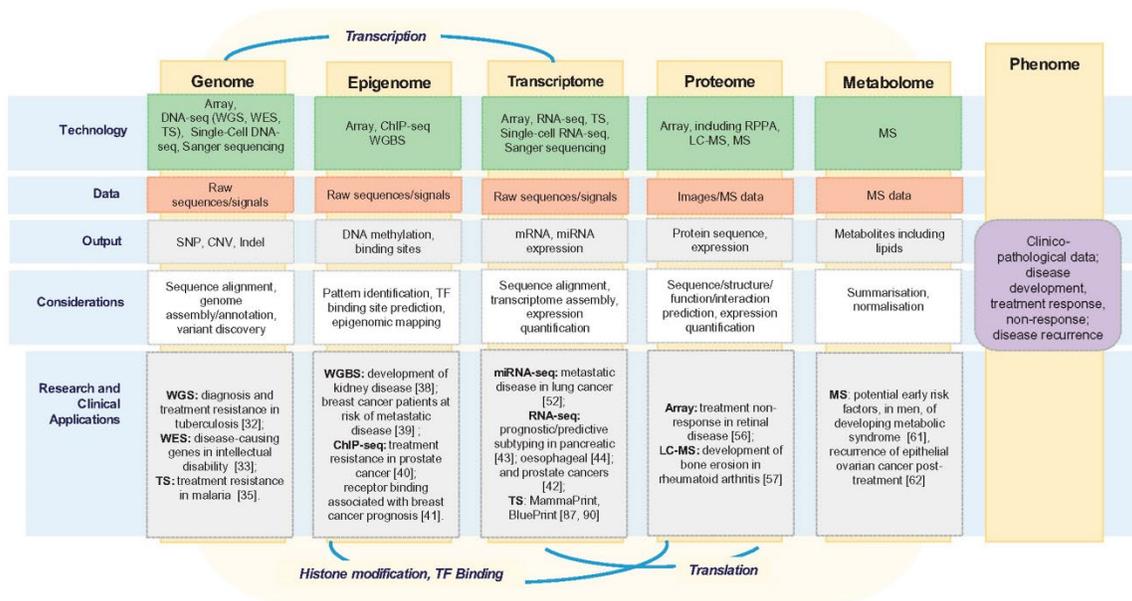

**Figure 2.** Schematic representation of five omics levels. The genome, epigenome, transcriptome, proteome, and metabolome are represented, with technologies, types of data, data outputs, and important applications summarized. These five levels together comprise the patient characteristics or phenome. Important connections between the different levels are also represented, including transcription, histone modification and transcription factor (TF)-binding and translation. Reproduced from [129]. Copyright, The Authors, 2018. Published by Oxford University Press. Distributed under the terms of the Creative Commons Attribution License (http://creativecommons.org/licenses/by/4.0/).

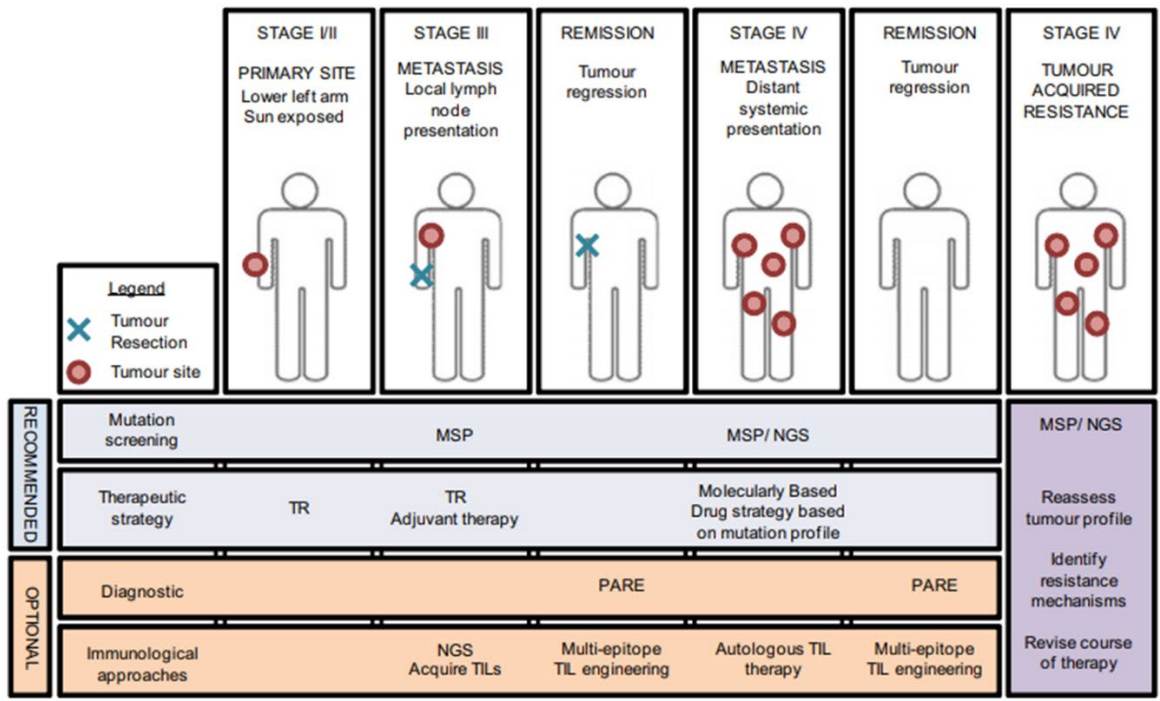

**Figure 3.** Mutational profiling in personalized therapy. Presentation of a patient with melanoma at different tumor stages. Included are recommended screening and therapeutic managements, which vary depending on tumor progression. NGS has become an important part of advanced melanoma care. MSP, melanoma-specific mutation panel; NGS, next-generation sequencing; TR, tumor resection; PARE, personalized analysis with rearranged ends; TIL, tumor infiltrating lymphocyte. Reproduced with permission from Ref. [103]. Copyright 2012, Elsevier.